\documentstyle[aps,epsf,multicol,pre]{revtex}
\begin{document}
\title{\bf Self-Organized Critical Random Boolean Networks} 

\author{Bartolo Luque, Fernando J. Ballesteros and Enrique M. Muro}  
\address{Centro de Astrobiolog\'{\i}a (CAB)  
INTA-CSIC, Ctra. de Ajalvir km. 4, 
28850 Torrej\'on de Ardoz, Madrid, Spain} 
\date{\today}
\maketitle

\begin{abstract}
Standard Random Boolean Networks display an order-disorder phase 
transition. We add to the standard Random Boolean Networks a
disconnection rule which couples the control and order
parameters. By this way, the system is driven to the critical line 
transition. Under the influence of perturbations the system point out
self-organized critical behavior. Several numerical simulations have
been done and compared with a proposed analytical treatment. 

\end{abstract}
{PACS numbers: 05.65.+b, 87.23.Ge, 87.23.Kg 

\begin{multicols}{2}

\section{Introduction}

Random Boolean Networks (RBN) were proposed \cite{Kauffman93} as
discrete genetic networks models.
The network is composed of nodes (automata).
The state of an automaton represents the state of a gene with two
possible values (on-off).
This state is the output of a Boolean function, which has
as inputs the output activity of some other genes.
The connectivity of the system and the bias used for the
Boolean functions are relevant parameters in order to statistically
determine the network dynamics.
If the system has a high connectivity  and a low bias, the dynamics of the 
automata is disordered; it seems that there is no correlation between
the gene switching on and off. On the other hand, the dynamics
is ordered if a low connectivity and a high bias are used. 
Only the ordered dynamics have biological sense.
S. Kauffman says that this parametric region offers ``{\it free
order}'' and
it seems that natural selection works where the order exist
previously \cite{Kauffman69}. In the critical region, which is the
boundary between ordered and disordered phase, there are some scaling
relations that have been the subject of recent works
\cite{Bhattacharjya96,Bastola98,Albert00}. S. Kauffman points out 
that genetic networks 
evolve to the boundary
between order and disorder. In this region there is more
diversity of patterns for activity and greater possibilities for 
complex evolution (anti-chaos hypothesis) \cite{KaufScAm}.

There is great interest in the evolution of topology in networks 
\cite{Barabasi,Dorogovtsev}. 
Several schemes have been proposed for RBN evolution.
For instance, 
in \cite{Bornholdt98} a RBN evolves from an initial $K=1$ mean 
connectivity (using a random
initial condition) to an attractor. At this point a copy 
of the network is made which has a connection randomly 
removed or/and added. The new network also reaches an attractor. If the
same attractor has been obtained, the network is maintained. 
Otherwise, the previous one is restored. During the RBN evolution,
it appears some stasi periods and punctuations 
like in real evolutionary processes.

In \cite{PaczuskiBC00} the performance of an automaton is
defined as the number of steps that it is in majority during a given
set of steps. An automaton in majority means that it has the same state
as the majority of the automata. The automaton
with the highest performance is replaced by another automaton
with a new random Boolean function and the process is repeated. 
The authors show how the genetic network is able to modify
its bias in order to reach the critical region.

In \cite{Bornholdt00} is presented a method which is 
able to lead the connectivity of binary neural networks to
their critical value. These neural networks have a phase transition
similar to RBN \cite{Kurten,Luque1}. 
Using a fixed connectivity initialization 
the network evolves to an attractor. If an automaton does not change
its state in the attractor, a connection is added to it.
Otherwise, if it changes then a connection is removed. 
The different networks (with different initial connectivities)
reach and remain close to the critical region. 

From a more general perspective, 
some works \cite{Maslov95,Paczuski96,Sornette96,Grassberger96,Sornette95}
have shown the relation between critical and self-organized
critical phenomena. In particular,  D. Sornette et al. \cite{Sornette95}
point out a heuristic method that transforms a system with a 
critical transition (Ising or bond percolation) into a self-organized
critical system. They propose to add some kind of mechanism (a
feedback between the order and control parameters) which slowly drives
the system towards the critical point.

RBN are a classical example where complex global behavior emerges from
local simple rules. They exhibit a phase transition similar to the
Ising model or bond percolation. Forgetting its initial biological
inspiration, our goal is to develop RBN which spontaneously evolve
towards a global critical stationary state. For this purpose we use a
disconnection rule which induces a feedback between the control and
order parameters. These networks reach a critical state without
changing externally the control parameter. These systems show
characteristics related to self-organized criticality.  Our evolution
method is distinct from earlier works in the following: we introduce a
well-defined coupling between order and control parameters as 
in \cite{Bornholdt00}, but the method is able to
stabilize (the individuals connectivities become constant in time)
the networks along the critical curve. 
 
In section II we give an introduction to RBN and we present 
the disconnection local rule which gives to the RBN a SOC behavior.
In section III statistical results of the evolution of the
self-organized RBN are presented. In section IV we
show its SOC behavior in response to external perturbations. In section V
we apply an analytical treatment to the model and discuss the results, 
and finally, in section VI we make a summary and point out for
future works.

\section{How does one make a critical self-organized RBN}

A RBN is a discrete dynamical system composed of $N$ automata. 
Each automaton is a Boolean variable with two possible states: 
$\{0,1\}$, such that
\begin{equation}
{\bf F}:\{0,1\}^N\mapsto \{0,1\}^N, 
\label{globalmap}
\end{equation} 
where ${\bf F}=(f_1,...,f_i,...,f_N)$ and each $f_i$ is a 
Boolean function of $K_i$ inputs (the automaton $i$ is connected to
$K_i$ automata randomly chosen from the set of $N$ automata):    
\begin{equation}
f_i:\{0,1\}^{K_i}\mapsto \{0,1\}.
\label{booleanfunction}
\end{equation}
An automaton state $ x_i^t \in \{0,1\}$ is updated using
its corresponding Boolean function:
\begin{equation}
x_i^{t+1} = f_i(x_{i_1}^t,x_{i_2}^t, ... ,x_{i_{K_i}}^t).
\label{update}
\end{equation}
Each $f_i$ is represented as a look-table of $K_i$
inputs. Initially, $K_i$ neighbors and
a look-table of bias $p$ are assigned to each automaton.
In order to generate such a look-table,
the value $1$ is assigned to an output with a probability $p$
and $0$ with a probability $1-p$.
When the neighborhood and the functions are established,
they are maintained (quenched).

We randomly initialize the states of the automata (initial 
condition of the RBN). The automata are updated synchronously using its
corresponding Boolean functions:
\begin{equation}
{\bf x}^{t+1} = {\bf F}({\bf x}^{t}),
\label{map}
\end{equation} 

These RBN exhibit a second order phase transition 
\cite{Derrida}. The control parameters ($K$ and $p$) 
determine two regions:
a frozen phase for $K < 1 /2 p (1-p)$ and a disordered phase for
$K > 1 /2 p (1-p)$. Thus, the critical boundary is represented 
as follows:
\begin{equation}
K_c(p) = {1 \over 2 p (1-p)}.
\label{kcritica}
\end{equation}

The above description corresponds to a classical RBN.
We incorporate a disconnection rule to the system leading the
RBN to a stationary critical state with connectivity $K_c(p)$ 
(see equation \ref{kcritica}). The rule is also applied 
synchronously to each automaton $i$ which has a local 
connectivity, $K_i(t)$, in such a way that:
\begin{enumerate}
\item 
Disconnection threshold: if $K_i(t)>2$ then 
\begin{eqnarray}
K_i(t+1)=K_i(t)-1, \;\;\mathrm{if} \sum_{j=1}^{K_i(t)}x_{i_j}<K_i(t)-1 \nonumber \\ 
K_i(t+1)=K_i(t), \;\;\mathrm{otherwise}. 
\label{break}
\end{eqnarray} 
\item
Minimal connectivity: if $K_i(t) =2$ then 
\begin{equation}
K_i(t+1)=K_i(t).
\label{quietorr}
\end{equation}
\end{enumerate}

The rule is inspired in the Bak-Tang-Wiesenfeld
(BTW) model \cite{BTW}. 
As in the sandpile-like model we have incorporated a threshold.
If the number of $0$'s in the input of a Boolean function
is greater than one and the local connectivity of its 
corresponding automaton is $K_i(t)> 2$, a connection of
this automaton (randomly chosen) will be cut. Then, a new 
Boolean function with 
connectivity $K_i(t)-1$ will be assigned.

The disconnection mechanism is illustrated 
in Figure \ref{algoritmo}.
The automaton $i$ has a connectivity $K_i(t)=3$ and an input
vector composed by two $0's$ and one $1$. Hence, a random connection
is cut ($K_i(t+1)=2$) and another Boolean function with bias $p=0.5$ is 
assigned. A new state is then computed.

In a sandpile-like system there is a decrease in the mass, energy,
tension, etc. at the boundaries \cite{BTW,SOC}.
In the proposed RBN, there is a
decrease in the number of connections. Our system is a random
graph with no boundaries. But, there is a diffusion 
effect which is able to cause an avalanche of disconnections.
If one automaton connection is cut (at $t$), its state can 
change from $1$ to $0$ (at $t+1$). The last change may
cause a disconnection (at $t+1$) in any of the automata that are connected to
the first one, and so on until a stationary state is reached.

\begin{figure}
\leavevmode
\epsfxsize=7.5 cm
\epsffile{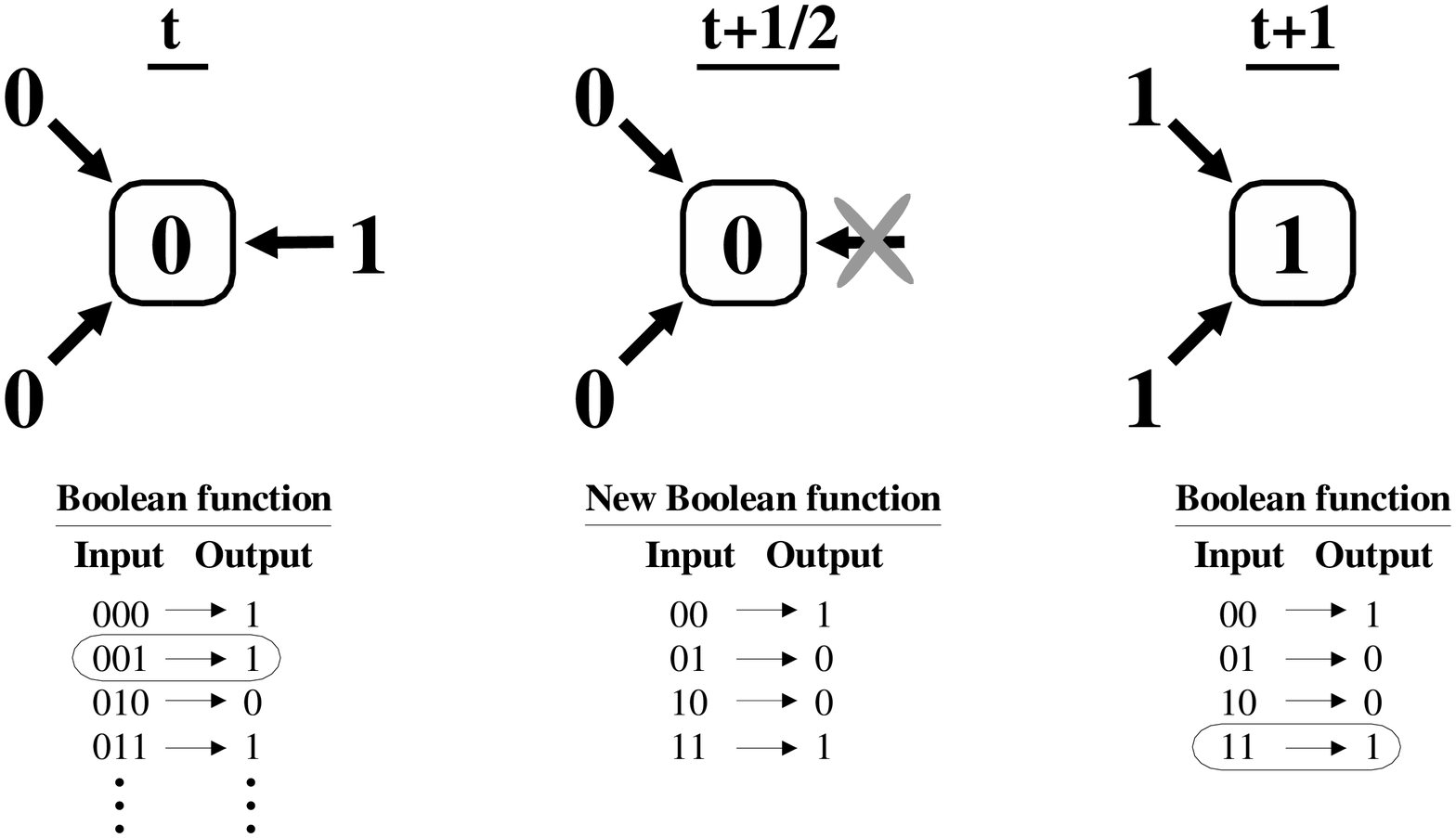}
\vspace{0.25 cm} 
\caption{Disconnection of an automaton $i$ with $K_{i}(t)=3$. The
automaton receives more than one $0$. At $t+1/2$ one 
of the automaton connections is randomly cut and its
Boolean function is changed (by a look-table with bias $p=0.5$).
The new Boolean function obtains a new state for the 
automaton in $t+1$.} 
\label{algoritmo}
\end{figure} 

\begin{figure}
\leavevmode
\epsfxsize=7.5 cm
\epsffile{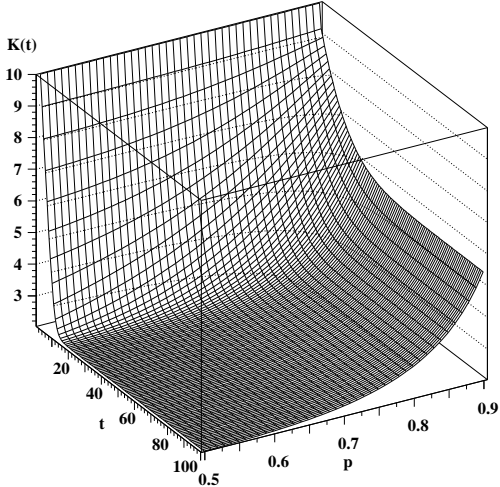} 
\caption{Average evolution of the mean connectivity $K(t)$,
where t are the iterations.
We have considered $1000$ RBN ($N=10000$): 
$100$ different RBN with 
$10$ initial conditions ($50\%$ of the automata with state $1$ 
and $50\%$ with state $0$ in $t=0$) for each value of the bias $p$. 
All the RBN have a  initial $K(t=0)=10$ mean connectivity.}
\label{k_media}
\end{figure}
\section{Results}

The average of the evolution of the mean connectivity $K(t)$ of 
$1000$ RBN, each one with $N=10000$ automata, 
is represented in Figure \ref{k_media}. 
Each RBN diminishes its connectivity due
to equations \ref{break} and \ref{quietorr}. A fixed bias $p$
has been used for every RBN: $100$ different RBN and for each one
$10$ initial conditions of the automata states 
($50\%$ of $1's$ and $50\%$ of $0's$ approximately). All the RBN
automata have a connectivity  $K_i(t=0)=10$.
The bias $p$ ranges from $0.5$ to $0.9$ in steps of $0.01$.
The stationary state (when no more connections are cut)
is quickly reached in about $50$ iterations. The mean connectivity 
$K(t)$ stabilizes close
to the transition curve between order and disorder ($K_c(t)$). 
This curve is determined by equation \ref{kcritica} and
it is represented as a continuous line in Figure  \ref{criticaline}.
In the same figure, the circles are the stationary result of the averaged 
mean connectivity (corresponding to section $p$ versus $K$ 
at $t=100$, in Figure \ref{k_media}).
It can be observed that the RBN stabilizes at the transition 
curve. In the same figure, three instances of
the evolution of three RBN ($p=0.50$, $p=0.70$ and $p=0.85$) have been
represented. The fill-down triangles represent the values of the RBN mean 
connectivity evolving towards equilibrium. In addition the
patterns evolution of the above three examples are represented in
Figure \ref{automatas}.  

\begin{figure}
\leavevmode
\epsfxsize=7.5 cm
\epsffile{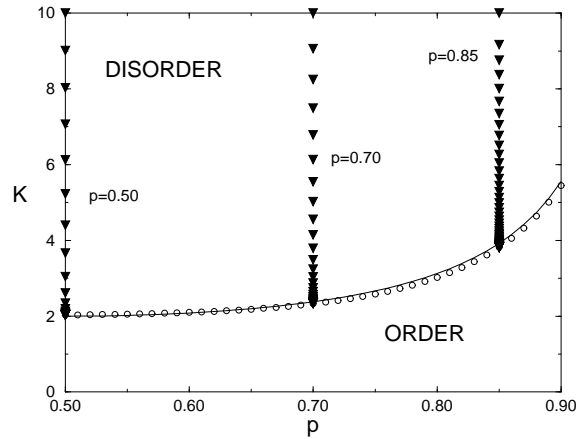} 
\caption{ (a) The continuous line represents the critical curve
obtained from the theory (equation \ref{kcritica}). It is a boundary
between the ordered and disordered phases of the RBN.  (b) The fill
down-triangles represent three instances of the RBN evolution for
$p=0.50$, $p=0.70$, and $p=0.85$.  Each triangle represents the mean
connectivity $K(t)$ in consecutive steps.  (c) The circles represent
the stationary state at the end of the RBN evolution ($K(t=0)=10$)
using different values for the bias $p$ (ranging from $p=0.5$ to
$p=0.9$ in $p$ increments of $0.01$). Each circle represents the
average of $1000$ RBN ($N=10000$) during $1000$ time-steps.}
\label{criticaline}
\end{figure}

The self-overlap $a(t)$ is the unitary percentage of automata
with the same value in $t-1$ and in $t$.
The stationary self-overlap, $a(t \to \infty) = a^*$, 
is an order parameter for the
RBN transition \cite{Luque}. Therefore in the disordered state we have
$a^*<1$
and in the ordered state the self-overlap is given by $a^*=1$. 
In Figure \ref{auto_a}
we have represented the self-overlap evolution (averages are
calculated
as in
 Figure \ref{k_media}).
The initial self-overlap is  $a(t=1)=0.5$ due to the
random state initialization. It can be observed that the
self-overlap grows quickly reaching the stationary value
$a^*=1$ for high values of the bias. 
This is to be expected because in the critical boundary with $K=K_c$, the
self-overlap reaches the value $a(t)=1$ for the first time. On the other
hand, for low values of the bias, the self-overlap is
stable at a value lower than $1$. We will show 
in section V that this is an RBN size effect.

In order to compare the different velocities to approach the stability
in the self-overlap and the connectivity, the self-overlap of the
connectivities $q(t)$ has been represented in Figure \ref{auto_k}. The
quantity $q(t)$ is defined as the unitary percentage of automata which
have the same connectivity at times $t-1$ and $t$. As it can be
observed, the connectivity is stabilized, approximately, in 50 time-steps. On
the other hand, the self-overlap needs about 100 time-steps to stabilize.

In short: the evolution of an RBN with dynamics given by
equations \ref{break}-\ref{quietorr} is such that the
RBN is driven from the disordered phase to the critical boundary 
(equation \ref{kcritica}) with mean connectivity
 $K_c(p)$ 
and self-overlap $a \approx 1$.

In the next Section we study the nature of the RBN equilibrium state,
and whether it is meta-stable or not.  In order to do so we analyse
the response of the system to external perturbations.
\begin{figure}
\leavevmode
\epsfxsize=7.5 cm
\epsffile{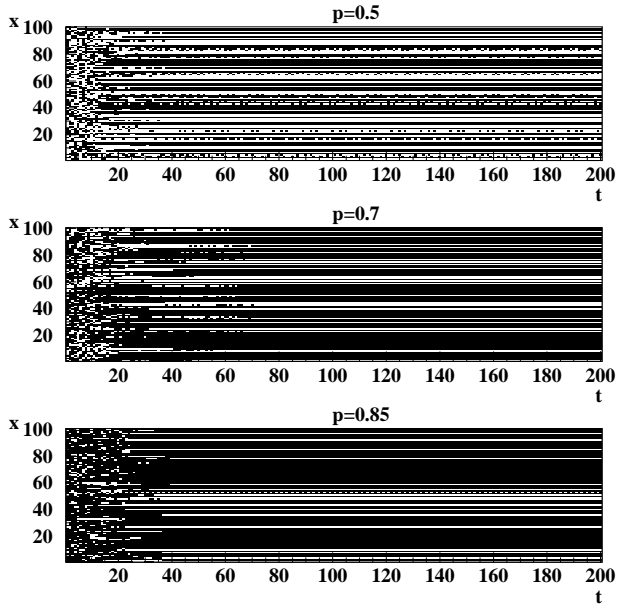} 
\caption{These figures show the spatio-temporal automata activity 
for the evolution of networks built of $N=100$ automata. Time runs
from left to right. 
The number of iterations is represented in the figure by $t$. 
The value of the automaton state (vertical axis)
is represented as black if it is one, and white otherwise.
From top to down: RBN with $p=0.50$, $p=0.70$,
and $p=0.85$.}
\label{automatas}
\end{figure}
\begin{figure}
\leavevmode
\epsfxsize=7.5 cm
\epsffile{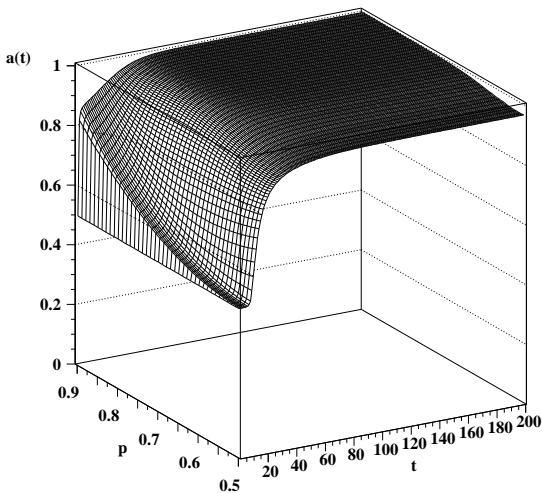} 
\caption{Evolution of the average self-overlap $a(t)$ 
for the simulations of Figure \ref{k_media} (t is the number of iterations).}
\label{auto_a} 
\end{figure} 
\begin{figure} 
\leavevmode
\epsfxsize=7.5 cm
\epsffile{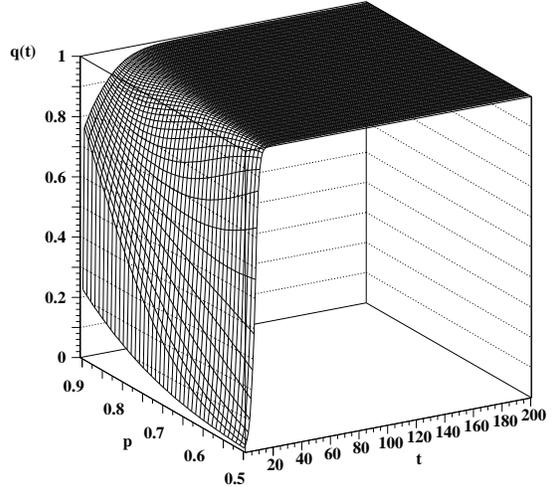} 
\caption{Average self-overlap of the connectivity $q(t)$
for the simulations in Figure \ref{k_media} (t is the number of iterations).} 
\label{auto_k}
\end{figure} 

\section{Perturbation Analysis}

We now evolve the RBN using the disconnection rule
(equations \ref{break} and \ref{quietorr}). We perturb
the RBN making use of 
sandpile-like methods \cite{BTW,SOC}. 
For this purpose we start with a relaxed RBN and a fixed bias $p$.
In order to perturb the RBN,  we randomly choose an
automaton $i$, and add to it a connection (i.e., its connectivity
changes from $K_i$ to $K_i+1$). A new Boolean function
$f_i$ with $K_i+1$ inputs is assigned to $i$ using the bias
$p$, so that one obtains a new state for the automaton $i$. 
If the automaton state is maintained, then there is 
only a small increase in the RBN mean connectivity.
If the automaton state changes, it is 
possible to cause a disconnection, originating an
avalanche via branching. When the avalanche stops, 
the RBN is minimally perturbed again, and so on.

In order to characterize the avalanches, we have measured
two different variables:
the time $T$ needed for the net to reach a
stationary state and the total number of disconnections $B$ during $T$. 
In Figure \ref{T} the histogram $S(T)$
of avalanche times is represented after $5\times 10^6$ perturbations. 
For this purpose we have used a RBN
with $p=0.65$ and different sizes:  $N=100$ (triangles), 
$N=1000$ (unfilled circles), $N=10000$ (filled circles), and 
$N=100000$ (diamonds). The dotted line is a visual guide 
to see the size effect.
It is easy to see that when $N \rightarrow \infty$, the histograms
tend to a power law.  
A fit with the first points gives 
$S(T) \sim T^z$ with $z= -2.9 \pm 0.1$.
Figure \ref{B} shows the histogram $S(B)$ of disconnections
using the same numerical simulations as before. 
We obtain again a possible power law described by
$S(B) \sim B^D$ with $D= -2.2 \pm 0.2$. In both figures, insets in
linear-log form are showed in order to exclude posible exponential fits.
\begin{figure} 
\leavevmode
\epsfxsize=7.5 cm
\epsffile{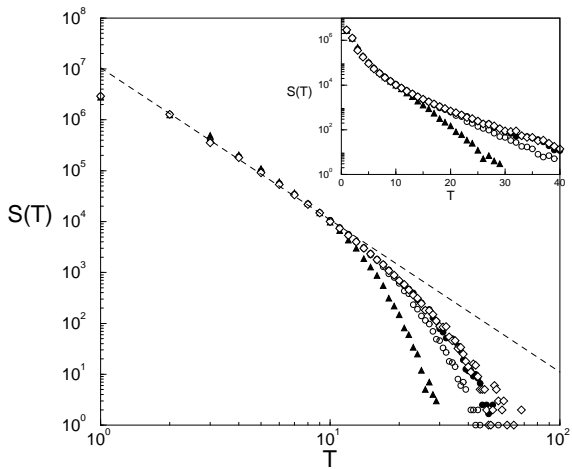} 
\caption{Log-log histogram ($S(T)$) for $T$ 
(iterations needed for the system to reach a stationary state)
during $5 \times 10^6$ perturbations
in a RBN.
 $p=0.65$ for $N=100000$ (diamonds), $N=10000$ (filled circles), $N=1000$ 
(unfilled circles), and $N=100$ (triangles). Inset: the same figure
but in linear-log form.}
\label{T}
\end{figure} 
\begin{figure} 
\leavevmode
\epsfxsize=7.5 cm
\epsffile{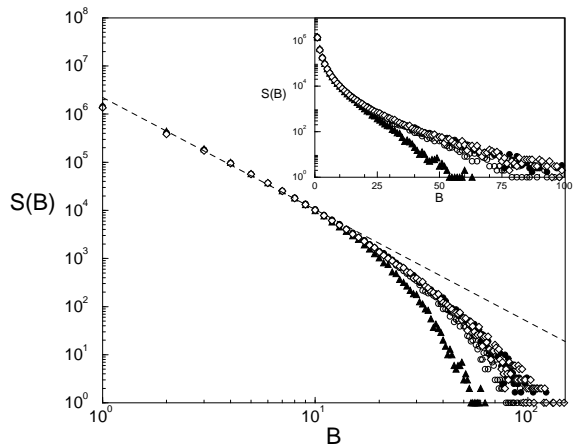} 
\caption{Log-log histogram for the sizes of the avalanches $B$.
$B$ is the number of disconnections for each perturbation and 
$S(B)$ is the histogram.
For this purpose it has been 
used the same simulations of Figure \ref{T}. Inset: the same figure
but in linear-log form.} 
\label{B}
\end{figure}
In the inset of Figure \ref{fourier} we show the
temporal evolution of the mean connectivity $K(t)$ for a 
perturbed RBN ($N=10000$, $p=0.65$). In Figure \ref{fourier}
we represent the power spectrum $S(f)$ (log-log) obtained by
averaging $305$ temporal series of $16384$ time-steps. 
A power law described by $S(f) \sim f^\phi$ with $\phi=-1.92 \pm 0.09$
is in good agreement with the data.

The numerical results do not seem definitive. The power law regime for
$T$ and $B$ lasts almost for a decade, but there is a clear finite
size effect. There are strong computational restrictions
for working with bigger RBN because the size 
of the avalanches is limited by the number of RBN automata. 
Nevertheless, in the $S(f)$ histogram there is
scaling for more than two decades. In conclusion, we think that 
the results point out a SOC behavior.
\begin{figure} 
\leavevmode
\epsfxsize=7.5 cm
\epsffile{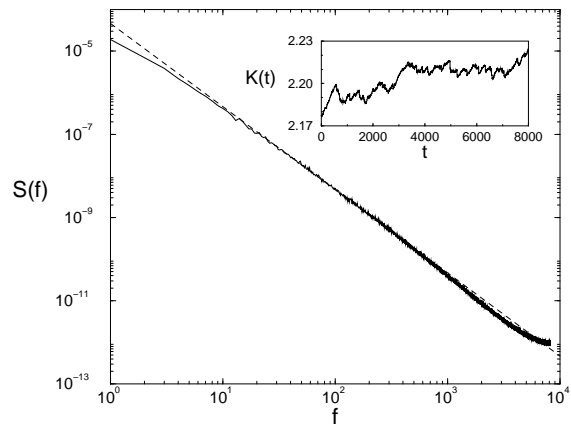} 
\caption{Power spectrum (log-log) of the mean connectivity evolution
$K(t)$ of a RBN ($p=0.65$ and 
$N=10000$). It is the average of $305$
temporal series of $16384$ steps. The slope of the dotted
line is $\phi=-1.92 \pm 0.09$.
Inset: a particular case of the evolution of the mean connectivity
$K(t)$ for the average series (t is the number of iterations).} 
\label{fourier}
\end{figure} 

\section{Analytic treatment}

At this point we ask ourselves:
why are the evolution rules able to drive an RBN to the
critical boundary between the order and disorder phases? 
As the bias has been already fixed, 
there can only be variations in the mean connectivity
$K(t)$ (acting as order parameter) and the self-overlap $a(t)$
(control parameter). In this section we show that
there is a feed-back mechanism between both parameters that leads the
system to a SOC behavior \cite{Sornette95}.

As we mentioned before, the self-overlap $a(t)$ is the 
unitary percent of automata which has the same state in
$t+1$ and $t$. If the mean connectivity $K(t)$ is known,
the value of the self-overlap at time $t+1$ can be determined by
the following equation:
\begin{equation}
a(t+1)=a^{K(t)}(t)+{\mathcal P} (1-a^{K(t)}(t)).
\label{a_evolution}
\end{equation}

If we interpret $a(t)$ as the probability for an arbitrary automaton to
remain in the same state at both $t-1$ and $t$, the term $a^{K(t)}(t)$
gives the probability that all the inputs of a given automaton are
the same for $t-1$ to $t$.  It is clear that $1-a^{K(t)}(t)$ is the
probability that at least one of the inputs will be different
at $t$ and $t-1$. In that case, there is still a probability
that this particular automaton 
remains in the same state at $t$ and $t+1$ by chance.
This probability is given by ${\mathcal P}=p^2+(1-p)^2$ \cite{Luque}. 
In a standard RBN the mean connectivity $K$ does not have an explicit
dependence on time. If one includes the disconnection rule described above, 
the mean connectivity of the system evolves with time. If $P_k(t)$ is 
the probability of an automaton to have connectivity $k$, then:
\begin{equation}
K(t)=\sum_{k=2}^{10} P_k(t)k,
\label{mediadistribucion}
\end{equation}
where the maximum connectivity value has been taken as $10$, without
loss of generality. The connectivity distribution evolves according to
the following system of equations:
\begin{mathletters}
\begin{eqnarray}
P_{10}(t+1)=[\Phi_{10}+(1-\Phi_{10})a^{10}(t)]P_{10}(t) 
\label{sistema1}
\\
\vdots 
\nonumber \\
P_{k}(t+1)=(1-\Phi_{k+1})(1-a^{k+1}(t))P_{k+1}(t) \nonumber \\
+[\Phi_{k}+(1-\Phi_{k})a^{k}(t)]P_{k}(t)
\label{sistema2}
\\
\vdots 
\nonumber \\
P_2(t+1)=(1-\Phi_{3})(1-a^{3}(t))P_{3}(t)+P_2(t),
\label{sistema3}
\end{eqnarray}
\end{mathletters}
where $k=9,8,...,3$. The variable $\Phi_k=p^k+kp^{k-1}(1-p)$ is the
automaton probability to maintain its connectivity by chance. Equation
\ref{sistema1} describes the loss of connections for all the automata
with connectivity $10$ at instant $t$.  The rest of the equations 
\ref{sistema2},
except for the last one \ref{sistema3}, have two contributions: the
first one represents the creation of automata with connectivity $k$
that previously had connectivity $k+1$, and the second one describes the
automata that maintain their same value of connectivity $k$. 
The last equation \ref{sistema3} describes the growth of automata population
with connectivity $2$.  The evolution of the critical self-organized
RBN is described by the coupled system of equations \ref{a_evolution}
and \ref{sistema1}-\ref{sistema3}.

One can see than $a^*=1$ is the only possible value that makes 
the probability distribution become stationary. In equation
(\ref{a_evolution}), $a^*=1$ is a fixed point when 
$K^* (1-{\mathcal P})\le 1$, where $K^*=K(t \to \infty)$ is the
asymptotic connectivity, that is, whenever the stationary mean
connectivity satisfies:
\begin{equation}
K^* \le {1 \over 2 p (1-p)},
\label{critica}
\end{equation}
whose boundary is the critical curve depicted in Figure \ref{criticaline}.
From the system of equations \ref{sistema1}-\ref{sistema3} 
it can be easily deduced that:
\begin{eqnarray}
K(t+1)=K(t)-(1-P_2(t))\nonumber\\
-\sum_{k=3}^{10}{[\Phi_{k}+(1-\Phi_{k})a^{k}(t)]P_{k}(t)},
\label{evokmedia}
\end{eqnarray}  
so that the mean connectivity of the system
always  decreases during its dynamical evolution.
Therefore, if the initial condition satisfies  $K(t=0) >1 /(2 p(1-p))$, 
it will slowly fall towards the critical condition.
The first time  $a^*=1$ is  when the system reaches the
critical curve and it, therefore, stabilizes.
\begin{figure}
\leavevmode
\epsfxsize=7.5 cm
\epsffile{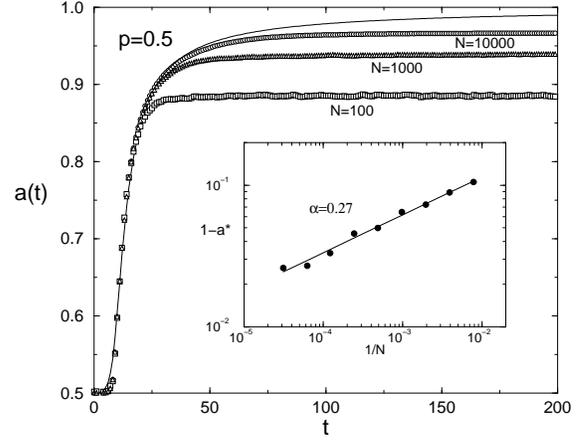} 
\caption{
The theoretical evolution of the self-overlap 
in a RBN.
Where t is the number of iterations.
$p=0.5$, $a(t=0)=0.5$, and $K(t=0)=10$.
All the automata have connectivity 10, i.e., $P_{10}(t=0)=1.0$
(continuous line).
The circles correspond to the numerical simulation for
 $N=10000$, the triangles for $N=1000$ 
and the squares for $N=100$. In the
inset, the filled circles represent the average stationary values
of the self-overlap reached by the RBN with different sizes, and 
logarithmically spaced out.}
\label{efectotama} 
\end{figure} 
We can study more rigorously the evolution of the system by the
classical analysis of linear stability. We perturb the fixed point
 $\Omega^* \equiv (a^*=1,P_{10}^*,...,P_k^*,...,P_2^*)$, with
$K^*=\sum_{k=2}^{10} P_k^*k$, and
compute the Jacobi matrix of the system at $\Omega^*$:
\begin{equation}
{\bf L}(\Omega^*) = \left ( 
\matrix{         K^*(1-{\mathcal P})  &   0    & \vdots  & 0     & \vdots &  0  \cr
10(1-\Phi_{10})P^*_{10}    &   1    & \vdots  & 0     & \vdots &  \vdots  \cr
	               \vdots    & \vdots & \ddots  &\vdots & \vdots &  \vdots  \cr
k(1-\Phi_{k})P^*_{k}       &        &         &       &        &          \cr
-(k+1)(1-\Phi_{k+1})P^*_{k+1}    & \vdots   & \vdots  &   1   & \vdots &  \vdots \cr
                 \vdots    & \vdots & \vdots  &\vdots & 1 &  0  \cr
                   -3(1-\Phi_{3})P^*_{3}       &    0   &  0      &    0  &   0    &  1}
\right ).
\end{equation}
We can calculate the characteristic polynomial by means of the
determinantal
condition:
\begin{equation}
\bigl |{\bf L}(\Omega^*)-\lambda {\bf I} \bigr |=0.
\end{equation}
Thus:
\begin{equation}
P(\lambda) = (K^*(1-{\mathcal P})-\lambda)(1-\lambda)^9= 0 ,
\end{equation}
and then the associated eigen-values are $\lambda_1=K^*(1-{\mathcal
P})$ and the (9 times) degenerated eigen-value $\lambda_2=1$. 
This means that 
for $\lambda_1 \le 1$ the critical point will indeed be linearly stable.
One can see that the characteristic polynomial is independent of the
stationary distribution  $\{P^*_k\}$. Obviously this distribution depends on
the initial conditions, but fulfills the condition
$K^*=\sum_{k=2}^{10}P_k^*k$ and thus the condition \ref{critica}.
\begin{figure}
\leavevmode
\epsfxsize=8.5 cm
\epsffile{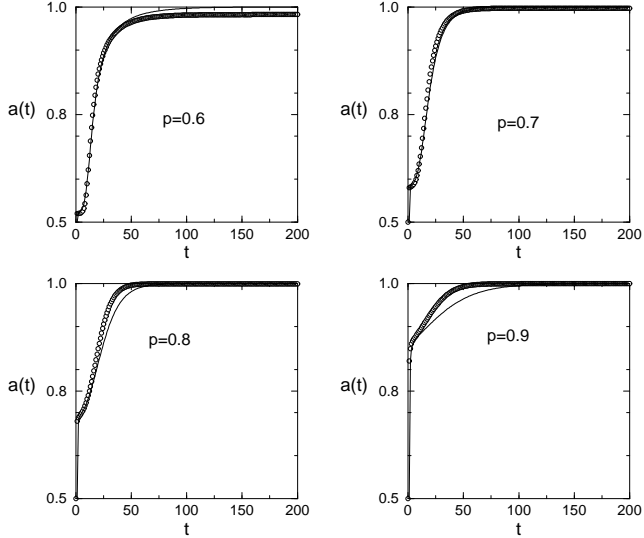} 
\caption{Theoretical evolution (continuous line) and numerical
simulation (circles) of  $a(t)$ for RBN ($N=10000$). We have performed $1000$
averaged simulations with four different values for the bias $p$
(t is the number of iterations).}
\label{asvarias} 
\end{figure} 
\begin{figure}
\leavevmode
\epsfxsize=7.5 cm
\epsffile{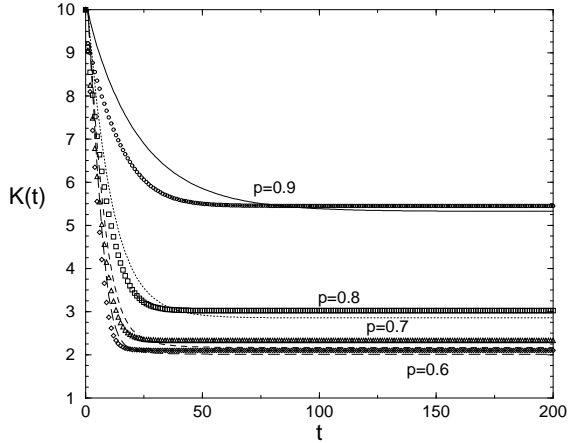} 
\caption{Evolution of the mean connectivity  $K(t)$ (t is the 
number of iterations): the lines
correspond to the theoretical results and the symbols to
the averaged simulations. We have considered the same cases as in 
Figure \ref{asvarias}.}
\label{evolucion} 
\end{figure} 
In order to check this theoretical analysis 
we have performed several numerical simulations.
In Figure \ref{efectotama} the theoretical 
evolution of the self-overlap $a(t)$ has been represented with continuous line.
We have used the following values: $a(t=0)=0.5$, $p=0.5$, and
$P_k(t=0)=0.0$ for all $k$ but $P_{10}=1.0$, that is, $K(t=0)=10$. It can be 
seen that the self-overlap converges asymptotically to $1$. 
With the same parameters and 
initial conditions we have performed our set of numerical simulations. 
In the same figure we have also plotted
the average evolution of the self-overlap for
$1000$ RBN of different sizes: $N=10000$ (circles), $N=1000$ 
(triangles), and $N=100$ (squares).
As can be seen from the plot, 
the first steps of the simulation agree with the 
theoretical curve. It can be observed that the stationary state of 
the simulations approaches the theoretical one as $N$ increases.
In order to perform a scaling size effect study
we have represented the stationary states ($1-a^*$) for different sizes
(see the inset in Figure \ref{efectotama}).
The fitting shows that 
$(1-a^*) \sim (1/N)^{\alpha}$ with $\alpha= 0.27 \pm 0.05$,
in such a way that in the thermodynamic limit ($N \to \infty$)
we get $a^* =1$, as  is to be expected.

In a similar way, the four graphs in Figure \ref{asvarias} show 
the theoretical
evolution (continuous line) and the numerical simulation (circles)
of the overlap for RBN ($N=10000$). For each graph we have used $1000$
averaged simulations and four different values for the bias $p$. In Figure 
\ref{evolucion} we have plotted the evolution
of the theoretical $K(t)$ mean connectivity (lines) and simulations
with RBN (symbols) for the same bias values of
Figure \ref{asvarias}.  Finally, in Figure \ref{fsubk} we  show $\{P_k^*\}$ 
the stationary distributions
as functions of the bias $p$: theoretical (top one ) and simulation (bottom).

\begin{figure}  
\leavevmode
\epsfxsize=7 cm
\epsffile{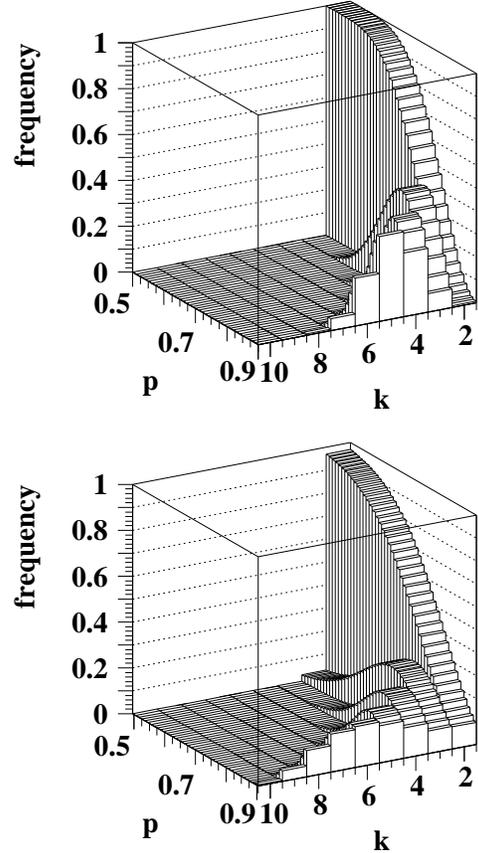} 
\caption{Stationary distribution of the connectivity $P^*(k)$ 
with $k=2,3,...,10$. In the bottom graph each distribution has been
calculated averaging over 1000 RBN. In the top one we show the
theoretical stationary
distribution of connectivities.}
\label{fsubk} 
\end{figure} 

From the previous figures it can be clearly observed that the 
 computer RBN simulations agree with our theoretical analysis, in spite of 
the size effect, that is due to computational restrictions.

\section{Summary}
 
RBN are classic complex systems which show that global order
is able to emerge from local rules. They exhibit a phase order-disorder 
transition modulated by the values of their connectivity and bias.
Some authors have already proposed mechanisms for RBN 
evolution \cite{Bornholdt98,PaczuskiBC00,Bornholdt00}. 
We have,  
following the Sornette et al. criterium \cite{Sornette95},
coupled the control parameter (connectivity for
a fixed bias) and the order parameter (self-overlap) in
order to convert a critical system into a self-organized critical one. 
From the numerical simulations it can be deduced that the
system evolves from the disordered phase to the critical
transition curve independently of the fixed bias. Moreover,
the response to external perturbations points to a SOC
behavior.

The theoretical analysis fits with the numerical simulations and it is
able to predict RBN behaviors to govern variations.  We tested other
disconnection rules (similar to \ref{break} and \ref{quietorr}): for
example, we change the breaking threshold of a connection. If there is
a disconnection when no $0$-inputs arrive to the automata, the system
stabilizes under the critical curve.  If there is a disconnection when
there are more than one $0$-inputs in the automata, the system
stabilizes over the critical curve.  In the theoretical analysis, this
modifications imply that the probabilities $\Phi_k$ change.  In the
first case the probabilities increase and in the second case they
decrease.  The values of these probabilities control the rate at which
the mean connectivity decreases. In the first case, the rate is
greater and the self-overlap reaches the value $1$ deep inside the
ordered phase. In the second case, the rate is very small and the net
stabilizes in the disordered phase with self-overlap lower than
$1$. There is a range for the values of the $\Phi_k$ functions that
gives the correct rate. The chosen rule produces values in that range.
From the theoretical point of view it seems easy to obtain similar
results using a fixed connectivity and changing the bias, or modifying
both of them simultaneously. 

Much future work is open. We think that the order-control coupling
parameter proposed in \cite{Bornholdt00} also works for RBN. And vice-versa: the
proposed method in this article works in this type of binary neural 
nets. In order to consider biological implications the automata (to be
considered as genes) dynamics and the disconnection rule (to be
thought of as mutations) should have different characteristic times.
We should also include a rule that allows for a possible connection
growth, and not simply consider connection loss as we have done in
this article.  We think that the point of view considered here can
have further applications along these lines.  In any case, we can
already extrapolate our results to a different type of nets that show
a clear transition between the ordered and disordered phases with well
defined order and control parameters.

\vspace{1 cm} 
 
{\bf Acknowledgments:}
The authors would like to thank C. Molina-Par\'{\i}s, U. Bastolla, 
and S. C. Manrubia and the referee for their critical reading of the manuscript. 
This work has been supported by Centro de Astrobiolog\'{\i}a (CAB).

\end{multicols}

\end{document}